\newcommand{\CLASSINPUTtoptextmargin}{1.9cm}
\newcommand{\CLASSINPUTbottomtextmargin}{2.57cm}
\documentclass[conference, usletter]{IEEEtran}

\IEEEoverridecommandlockouts
\usepackage{cite}
\usepackage{amsmath,amssymb,amsfonts}
\usepackage{algorithmic}
\usepackage{graphicx}
\usepackage{textcomp}
\usepackage{xcolor}
\usepackage{array} 
\usepackage{booktabs}
\usepackage{hyperref}
\usepackage{multirow} 
\usepackage{float}
\usepackage{subfig}

\def\BibTeX{{\rm B\kern-.05em{\sc i\kern-.025em b}\kern-.08em
    T\kern-.1667em\lower.7ex\hbox{E}\kern-.125emX}}
    
\begin{document}

\title{ECHO: Environmental Sound Classification with Hierarchical Ontology-guided Semi-Supervised Learning\\
}

\author{
    \IEEEauthorblockA{
        Pranav Gupta\IEEEauthorrefmark{1}, 
        Raunak Sharma\IEEEauthorrefmark{1}, 
        Rashmi Kumari\IEEEauthorrefmark{1}, 
        Sri Krishna Aditya\IEEEauthorrefmark{1},
        Shwetank Choudhary\IEEEauthorrefmark{2}, 
        Sumit Kumar\IEEEauthorrefmark{2}, \\
        Kanchana M\IEEEauthorrefmark{1}, 
        Thilagavathy R\IEEEauthorrefmark{1}\\
        \IEEEauthorrefmark{1}School of Computing, SRM Institute of Science and Technology, Kattankulathur, Tamil Nadu, India\\
        Emails: pm4043@srmist.edu.in, rr5913@srmist.edu.in, rd5473@srmist.edu.in, sr5160@srmist.edu.in, \\kanchanm@srmist.edu.in, thilagar2@srmist.edu.in
    }\\
    \IEEEauthorblockA{
        \IEEEauthorrefmark{2}Samsung Research Institute, Bengaluru, India\\
        Emails: sj.choudhary@samsung.com, sumit.kr@samsung.com
    }
}

\maketitle

\begin{abstract}
Environment Sound Classification has been a well-studied research problem in the field of signal processing and up till now more focus has been laid on fully supervised approaches. Over the last few years, focus has moved towards semi-supervised methods  which concentrate on the utilization of unlabeled data, and self-supervised methods which learn the intermediate representation through pretext task or contrastive learning. 
However, both approaches require a vast amount of unlabelled data to improve performance. In this work, we propose a novel framework called Environmental Sound Classification with Hierarchical Ontology-guided semi-supervised Learning (ECHO) that utilizes label ontology-based hierarchy to learn semantic representation by defining a novel pretext task. In the pretext task, the model tries to predict coarse labels defined by the Large Language Model (LLM) based on ground truth label ontology. The trained model is further fine-tuned in a supervised way to predict the actual task. Our proposed novel semi-supervised framework achieves an accuracy improvement in the range of 1\% to 8\% over baseline systems across three datasets namely UrbanSound8K, ESC-10, and ESC-50.

\end{abstract}


\section{Introduction}
\label{sec:introduction}
Due to advancements in deep learning, Environment Sound Classification (ESC) \cite{Tripathi2021-nb,Zhang_undated-op, Tokozume_undated-ph} along with speech recognition has seen greater interest from researchers. ESC typically involves detecting various indoor and outdoor sounds such as vehicle horns, winds, electronic instruments sound, etc. Detecting these sounds correctly enables its utility in several potential areas such as urban noise detection, health care monitoring, and smart home systems. However, ESC is more challenging compared to other common recognition tasks such as music classification, speech recognition, etc. mainly due to unstructured data and lower Signal-to-Noise Ratio (SNR)\cite{Tripathi2021-nb}.

To detect ESC task, traditional methods employ
shallow learning algorithms such as SVM, Random forest, etc., and uses statistical features such as Mel Frequency Cepstral Coefficients (MFCC), Gammatone and Chroma features. However, they show low robustness in a high SNR environment and achieve limited accuracy. Recently, the success of deep learning in the vision domain has seen its adoption in the sound field as well. Most of the deep learning methods use log-mel spectrograms as input features and are based on Convolutional Neural Networks (CNN) \cite{boddapati2017classifying, arora2017study}, Recurrent Neural Networks (RNN), or a combination of both to achieve better performance.

While deep learning models are able to perform better, as they get deeper and more complicated, they need large annotated datasets to achieve optimal performance\cite{shin2021, Tripathi2021-nb}. Obtaining a huge amount of annotated data is a cumbersome task and may require larger human efforts. More recently, to reduce dependency on the huge annotated datasets, several semi-supervised and self-supervised learning-based methods have been proposed in the vision domain \cite{chen2020big} and have been adopted in audio classification as well \cite{Tripathi2021-nb,shin2021}. Self-supervised learning-based algorithms leverage the vast amount of unlabelled data to learn task-agnostic semantic representations which are then used in downstream tasks through transfer learning. These methods learn representation by either defining novel pretext task \cite{Tripathi2021-nb} or by contrastive learning approach \cite{shin2021}. Similarly, semi-supervised learning methods also utilize unlabelled data or missing labels to improve the system's performance. They use techniques such as label enhancement, weak label prediction as ground truth, and mean teacher model \cite{guru2021} to achieve the desired performance.

Both semi and self-supervised learning methods require large unlabelled data to match the performance of a fully supervised system. In this work, we propose a novel semi-supervised learning-based framework that utilizes label ontology-based hierarchy to learn semantic representation by defining a novel pretext task. The proposed framework is different from existing work as it does not require any additional unlabelled dataset and instead utilizes implicit relations that existing labels share.
Implicit relations can be of various types such as semantic similarity, class similarity (such as dog, cat as animal), group similarity (such as traffic as outdoor and TV sound as indoor), etc. We use LLM-based prompt engineering to derive new semantic labels (see Table \ref{tab:prompts_result}) based on existing label ontology. The size of newly generated labels is always less than the existing label size. Then, we define the pretext task as the prediction of newly generated labels, and this phase is called coarse learning. During the coarse learning phase, the model tries to learn high-level parent classification (or coarse label prediction) and gets penalized whenever it fails to predict a semantically similar class. This enables the model to learn a semantic representation of similar classes which helps improve model performance in the main task.
In the next stage, we adopt the model from the coarse stage and fine-tune it on the actual task in a supervised way. We see an improvement in the range of 1\% to 8\% in accuracy over the baseline system for three benchmark datasets namely UrbanSound8K\cite{Salamon2014-rb}, ESC-10\cite {Piczak2015-jy} and ESC-50\cite {Piczak2015-jy}.

Our contributions can be summarized in the following way:

1. We propose a novel semi-supervised learning-based framework called as ECHO (Environmental Sound Classification with Hierarchical Ontology-guided semi-supervised Learning) that utilizes label ontology-based hierarchy to learn semantic representation by defining a novel pretext task. The trained model is used as transfer learning for the main task, which improves performance over baseline.

2. We propose to use LLM-based prompt engineering to automatically generate new labels for pretext tasks that are based on the existing labels ontology.

3. We provide a detailed analysis of the impact of choosing different sizes of parent labels for the pretext task.

4. We present a visualization of embeddings learned by our proposed framework using t-SNE graphs.

\section{Related Work}
\label{sec:related_work}
\subsection{Environment Sound Classification}
Environmental sound classification (ESC) is an emerging field within machine learning, which aims to identify everyday sounds for a multitude of applications. Recent advancements in deep learning for image processing have attracted researchers to adopt these methods for sound classification. Raw audio files can be represented as 2D log-mel spectrogram in frequency domain data and hence CNN-based popular architecture can be adopted for sound classification.\cite{arora2017study} discusses transfer learning using a pre-trained network for ESC tasks which can help achieve optimal performance. Work by \cite{li2018ensemble} demonstrates an ensemble of deep learning models can achieve better performance. All these
works have used a 2D log-mel spectrogram for ESC tasks. We, in this work also adopted log-mel as input features for our proposed ECHO framework. Slightly different work proposes a 1D CNN-based end-to-end method for ESC which takes raw waveform as input and does not require any pre-processing. \cite{khamparia2019sound} proposed to fine-tune popular image-based pre-trained architecture such as AlexNet, and GoogleNet for ESC tasks.

\subsection{Self-Supervised Learning}
Training deeper and more complex models that have achieved state-of-the-art performance requires a huge amount of annotated data which can be labor intensive and comes with acquiring cost. More recently, researchers have proposed to use self-supervised learning (SSL) to learn meaningful representations from unlabelled data and use this representation for downstream tasks. They have been shown to achieve the performance of fully supervised tasks with a fraction of data and even surpass in some
cases. SSL aims to leverage the vast amount of unlabeled data by learning intermediate representations by either 1) defining pretext tasks whose weights are
further used for transfer learning or 
2) Using contrastive learning. 

\cite{gidaris2018unsupervised} defined pretext task as predicting the angles of images, which were randomly rotated in advance.
In \cite{zhang2016colorful}, a given single channel is used to predict the other two channels for RGB images. Very few works are available in SSL for sound detection. \cite{Tripathi2021-nb} is one of the works that is more relevant to our proposed framework. They define pretext task as the identification of the type of data augmentation applied to the signal and weights are further fine-tuned for ESC. They use ResNet-18 architecture for experimentation and demonstrate their results on ESC-10. SSL works which are based on contrastive learning use data augmentation to learn the similarity between augmented samples of the same image. \cite{bachman2019learning} proposes a scheme to maximize the mutual information in the features extracted from cropped images from multiple locations of an original image. In the audio domain, \cite{tagliasacchi2020pre} proposed contrastive learning-based SSL using three different techniques Word2Vec, Audio2Vec, and temporal gap. In our proposed ECHO framework, we propose a novel pre-text based on an implicit hierarchy of existing labels and solve it to learn a semantic representation of similar classes.

\subsection{Semi-Supervised Learning}
Similar to SSL, semi-supervised learning (SmSL) also utilizes large-scale unlabeled data along with labeled data. One of the most common approaches in SmSL is to first train on limited labeled data and generate weak labels on unlabeled data. These weak labels are then considered as ground truth for unlabeled data and both datasets are used for further training. \cite{chen2020big} proposes a SmSL scheme that uses a pre-trained teacher network fine-tuned on a small fraction of samples. In the audio domain, \cite{guru2021} is one of the recent works using SmSL for audio classification. They propose label enhancement by a two-stage teacher-student learning process. In another approach, the same
work uses mean teacher-based SmSL to achieve optimal performance on audio classification.

In this work, we adopt the concept of pre-text tasks from SSL-based methods. However, our proposed ECHO framework is different from the existing work as it does not require any unlabeled data. Instead, we use the existing labeled dataset to derive parent labels using label ontology knowledge, and hence our method fits more suitable under a semi-supervised regime.

\begin{table}[b]
\addtolength{\tabcolsep}{-7pt}
\caption{Label Ontology derive using prompt engineering on LLM}
\label{tab:prompts_result}
\centering
\begin{tabular}{|l|c|c|}
\toprule
& \multicolumn{1}{c|}{UrbanSound8K} & \multicolumn{1}{c|}{ESC-10} \\ 
\midrule
\multirow{2}{*}{$p=2$}    &  Human-Animal Activities    & Natural Sounds \\
                          &  Mechanical-Environmental Noises    & Human-Man-Made Sounds \\
\midrule

\multirow{2}{*}{$p=3$($\sqrt{n}$)}          & Machinery-Construction Sounds & Nature-Animals \\
                                             & Emergency -Alerts       & Mechanical Sounds \\
                                             & Urban- Street Life     & Human-Related Sounds \\
\midrule

\multirow{2}{*}{$p=5$}           & Construction- Maintenance       & Natural Elements \\
                                 & Transportation- Vehicles     &  Mechanical-Man-made Sounds \\
                                 & Emergency-Alerts        & Human-Generated Sounds \\
                                 & Urban Leisure        & Animal Sounds \\
                                 & Residential-Animal Sounds     & Water-related Sounds \\
\bottomrule
\end{tabular}
\end{table}
\section{Dataset}
\label{sec:dataset}
We train and test our proposed approach on several benchmark datasets such as UrbanSound8K \cite{Salamon2014-rb}, ESC-10 \cite {Piczak2015-jy}, and ESC-50 \cite {Piczak2015-jy}. 

1). UrbanSound8K:  It consists of a total of 8732 recordings belonging to 10 classes from urban sounds. Each clip has length $<$=4s. The dataset provides an official split into ten-fold. For all our experiments, we report ten-fold cross-validation accuracy for proper benchmarking.

2). ESC-50 \& ESC-10: The ESC-50 dataset contains 2000 recordings belonging to 50 classes and having a length of 5s. It consists of environmental sounds ranging from sounds of chirping birds to car horns and traffic sounds ESC-10 is a subset of the larger ESC dataset. 

Both ESC-50 and ESC-10 provide an official split into five folds. For all our experiments, we report five-fold cross-validation accuracy for benchmarking.

\section{Methodology}
\label{sec:methodology}
Figure \ref{fig:proposed_pipeline} demonstrates the overall design of our proposed ECHO framework. This section discusses data pre-processing, the proposed two-phase framework, and various baseline model architectures.

\subsection{Data Preprocessing}
\label{subsec:data_preprocessing}
For all the experimentation in this paper, we fix the sampling rate to 16kHz and downsample raw audio clips wherever required. To get the features from raw audio waveform, we convert all the audio clips to log-mel spectrogram using the Librosa library. Log-mel spectrogram represents frequency domain data and is one of the most common representations used in ESC tasks. We resize spectrograms to a fixed size of $224 \times 224$ pixels to feed to our baseline models which are discussed in detail in section \ref{subsec:baseline_arch}. Additionally, we replicate the spectrogram three times to simulate a three-channel image format as input to the models requires RGB images.

\subsection{Label Ontology}
\label{subsec:Label Ontology}
As discussed in section \ref{sec:introduction}, our pretext task is defined as the prediction of coarse labels derived from label ontology knowledge.
Only a few of the datasets such as AudioSet \cite{audioset_dataset}, FSD50K \cite{fsd50k2022}, ESC-50 \cite{Salamon2014-rb} provide label ontology explicitly whereas it is missing for other datasets.
This brings dependency on the dataset to provide label ontology and might limit the application of our proposed framework. To alleviate this limitation, we propose a novel and generic approach to get ontology or coarse categorization of given labels (based on implicit label similarity) by using LLM.

We perform prompt engineering to create desired prompts and feed them to the LLM to provide relevant coarse labels based on the similarity of original labels. 

The following is our prompt template to derive label ontology:
\begin{verbatim}
Dataset Classes:
- Class 1
- ...
- Class N
Number of Classes in Dataset = N
Number of Parent Classes to Generate = P
The task is to create P parent 
classes by identifying and leveraging the 
similarities across the N classes in the 
dataset.
\end{verbatim}
We use ChatGPT-4\cite{chatgpt4} for all our experiments to generate label ontology. Table \ref{tab:prompts_result} shows the results of the LLM output for different values of parent classes.

\begin{figure*}[h]
\centerline{\includegraphics[scale=0.7]{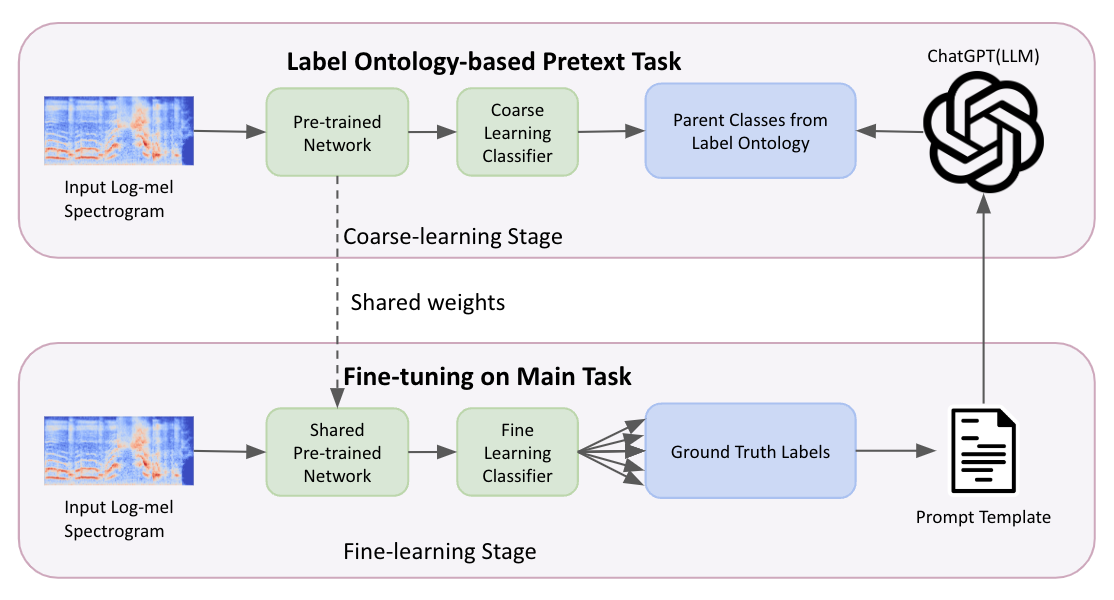}}
\caption{End-to-End Proposed Framework for Environment Sound Classification}
\label{fig:proposed_pipeline}
\end{figure*}
\subsection{Pretext Task for Coarse-Learning}
\label{subsec:pretext_task}
As discussed in section \ref{subsec:Label Ontology}, we use the LLM to automatically generate new labels based on label ontology knowledge of existing labels (classes). These newly generated labels are called coarse labels or parent labels.
Our pretext task is then defined as the prediction of the coarse labels. The number of classes in coarse labels is always less than the count of existing ground truth labels. We experimented with different numbers of coarse labels to be generated for various datasets which are discussed in detail in section \ref{subsec:ablation_study}. Each ground truth label (original class) belongs to at most one of the coarse labels or parent classes defined by LLM output as shown in Table \ref{tab:prompts_result}. To generate ground truth for the pretext task, each training sample is assigned one of the coarse labels to which its original label (or class) belongs. For example, if the original task is to predict one of the four sounds as TV, fan, traffic, or car horn. And say the LLM assigns TV, fan as indoor sounds and car horn, traffic as outdoor sounds based on label ontology. Then, the new ground truth for TV and fan data samples will be indoor sounds, and the same for car horns and traffic will be outdoor sounds.

Our motivation behind defining novel pretext tasks based on label ontology knowledge is;
1) Original ground truth labels might share semantic similarity owing to implicit relations between labels. This additional knowledge can be leveraged during the pre-training stage to learn meaningful representations which can be a starting point for the main task. 
2) The pretext task is the coarse learning stage where a model is penalized if it fails to correctly predict parent classes which is not the case when the main task is directly learned in all classes in a fully supervised way.
3) By focusing on these broader categories, the model learns to recognize overarching patterns that are intrinsic to groups of similar classes which helps the model to avoid getting confused between classes of similar nature during main task training.

Figure \ref{fig:proposed_pipeline} shows the framework of the pretext task. It consists of a pre-trained model and classifier for the pretext task. During the coarse stage, a pre-trained model is fine-tuned on the training dataset to learn meaningful representations. In the next stage, the classifier head is discarded and only fine-tuned model weights are adopted for the main task.

\subsection{Main task for fine-learning}
\label{subsec:main_task}
After the pre-training stage, we further fine-tune the trained network obtained after coarse learning on the main task (i.e. on the original classes). We call this stage fine learning and the end-to-end framework as coarse-to-fine learning. The fine learning stage involves a fully supervised way to train the network on the entire training dataset. It consists of trained weights obtained from the coarse stage along with the classifier head.

\subsection{Baseline architecture}
\label{subsec:baseline_arch}
For our proposed framework, we experiment with two variants of pre-trained ResNet architectures namely ResNet-50\cite{He2015-dz} \& ResNet-18\cite{He2015-dz} and two variants of pre-trained EfficientNet architectures namely EfficientNet-B0\cite{Tan2019-jw} \& EfficientNet-B1\cite{Tan2019-jw}. ResNet is a popular architecture based on residual connections that have shown great performance across vision domain tasks and have been widely adopted for ESC tasks. Similarly, EfficientNet is a lightweight architecture, popularly adopted in various vision domain problems.

For both our pretext and main task across datasets, we fix the classifier head consisting of two hidden layers with 512 and 256 neurons followed by a softmax-based classification layer. In the coarse stage, the defined pre-trained architecture with a classification head is fine-tuned end-to-end on the pretext task. Then, only the trained weights from the previous stage are fine-tuned for the main task.

\section{Results \& Discussion}
\label{sec:results_discussions}

\paragraph{Implementation details} We use the PyTorch framework for experimentation and model development. All training and testing are carried out on an NVIDIA GPU P100.

\begin{figure*}[ht]
	\centering
		\subfloat[]{\includegraphics[width=0.4\linewidth]{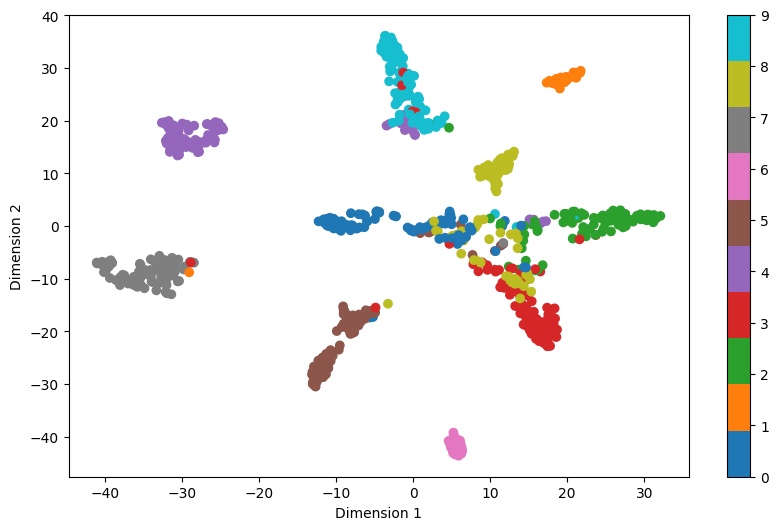}
		\label{fig:tsne_baseline}}
		\subfloat[]{\includegraphics[width=0.4\linewidth]{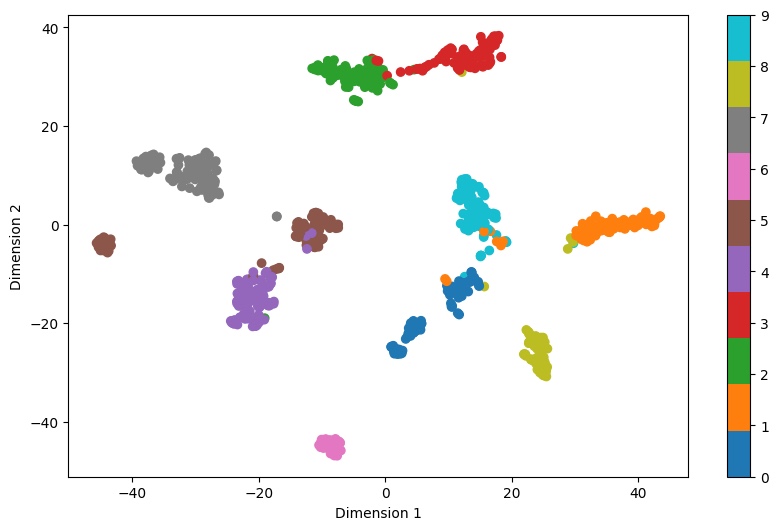}
		\label{fig:tsne_proposed}}
	\caption{Plots show t-SNE visualization of latent features obtained from the baseline model and proposed framework (ECHO on UrbanSound8K dataset a) ResNet-18 Baseline b) ResNet-18 with ECHO framework}
	\label{fig:curves}
	
\end{figure*}

\textbf{Hyperparameters}:
Every model takes a complete log-mel spectrogram as input of size 224 * 224 * 3 and is stacked 3 times, to create an RGB channel. We keep the batch size to 32 and use the Adam optimizer with a learning rate of 0.0001 across all experiments. All training is performed for 50 epochs and the best model is saved based on minimum validation loss. However, we observe that the model converges much before 50 epochs for most experiments. For the classifier head used in the framework, ReLU is used as the default activation function for intermediate layers and softmax for the classification layer.

\textbf{Loss Function}: Since our problem statement is a multi-classification problem, we use standard Cross-Entropy loss defined as 

\[
H(y, \hat{y}) = -\frac{1}{N} \sum_{i=1}^{N} \sum_{j=1}^{C} y_{ij} \log(\hat{y_{ij}})
\]

Where $H(y, \hat{y})$ is the Cross-Entropy Function, $y_{ij}$ is a binary indicator of whether class $j$ is the correct classification for observation $i$. $\hat{y_{ij}}$ is the predicted probability that observation $i$ is of class $j$. $N$ is the number of samples and $C$ is the number of classes.
 
\begin{table}[h]
\addtolength{\tabcolsep}{-2pt}
\caption{Summary of model performances on ESC-10}
\label{tab:model_performance}
\centering
\begin{tabular}{lcc}
\toprule
Model                & Accuracy (\%) & Classifier       \\
\midrule
PiczakCNN \cite{Piczak2015-jy}  & 80.50         & CNN              \\
GoogleNet \cite{boddapati2017classifying}  & 63.20         & CNN              \\
BC Learning \cite{Tokozume_undated-ph} & 91.30       & EnevNet          \\
Zhang et al. \cite{Zhang_undated-op} & 91.70       & CNN              \\
Triplet loss \cite{ssl-tripletloss} & 90.28       & Residual Network \\
Temporal GAP \cite{tagliasacchi2020pre} & 90.53       & Residual Network \\
Tripathi et al.\cite{Tripathi2021-nb} & 91.70 & SSL + Residual Network\\
\midrule
ECHO (ResNet-50)  & 96.00     & Semi-supervised \\
ECHO (EfficientNet-B1) & 97.50       & Semi-supervised \\
ECHO (ResNet-18)  & 95.75     & Semi-supervised \\
ECHO (EfficientNet-B0)  & 94.50     & Semi-supervised \\
\bottomrule
\end{tabular}
\end{table}

\subsection{Comparison with previous works}
\label{subsec:comparison_previous}

On the ESC-10 dataset, we provide a detailed comparison of our proposed framework with existing works shown in Table \ref{tab:model_performance}. Our proposed approach with both pre-trained models ( Resnet and Efficientnet ) as backbone surpasses all existing works
 performance. Among all, \cite{Tripathi2021-nb} is more relevant for fair comparison as they also use a ResNet-based architecture and employ
 a self-supervised learning strategy that involves training on data-augmented spectrogram signals through a pretext task. Our proposed technique significantly surpasses this baseline work by $>$4\% in accuracy. 
 
\begin{table}[t]
\addtolength{\tabcolsep}{-4pt}
\caption{Comparison of baseline with proposed ECHO framework}
\label{tab:baseline_proposed}
\centering
\begin{tabular}{lcccccc}
\toprule
\multirow{2}{*}{Model} & \multicolumn{2}{c}{ESC-50} & \multicolumn{2}{c}{ESC-10} & \multicolumn{2}{c}{US8K} \\
\cmidrule(lr){2-3} \cmidrule(lr){4-5} \cmidrule(lr){6-7}
                       & Baseline & Proposed & Baseline & Proposed & Baseline & Proposed \\
\midrule
ResNet-50               & 76.35       & \textbf{84.20}       & 91.25       & \textbf{96.00}       & 80.57       & \textbf{84.35}       \\
ResNet-18               & 80.75       & \textbf{81.75}       & 93.25       & \textbf{95.75}       & 80.02       & \textbf{84.22}       \\
EfficientNet-B0           & 77.55       & \textbf{82.50}       & 91.25       & \textbf{94.50}       & 82.36       & \textbf{86.97}       \\
EfficientNet-B1           & 80.10       & \textbf{86.05}       & 93.75       & \textbf{97.50}       & 80.93       & \textbf{87.66}       \\
\bottomrule
\end{tabular}

\end{table}

\subsection{Comparison of Baseline Architecture and Proposed Framework}
\label{subsec:comparison_baseline}
As shown in Table \ref{tab:baseline_proposed}, we also provide a comparison of our proposed framework with four pre-trained models adopted as baseline architectures.
For defining the baseline model as discussed in Table \ref{tab:baseline_proposed}, we use pre-trained weights of the provided backbone along with a classifier consisting of three layers (two dense layers of 512 and 256 neurons followed by a classification layer). This network is fine-tuned on the actual task in a fully supervised way.
We observe performance improvement due to our proposed framework in the range of 1\% to 8\%. We observe improvement across all baselines and all three mentioned datasets. This highlights the fact that intermediate representation learned by the pretext task helps improve the performance of the main task. Results show that our proposed framework is generic and can be adapted with various other backbones to improve performance. 

\begin{table}[b]
\caption{Ablation Study for the pretext task size}
\addtolength{\tabcolsep}{-5pt}
\label{tab:ablation_study}
\centering
\begin{tabular}{lcccccccccc}
\toprule
& & \multicolumn{3}{c}{ESC-50} & \multicolumn{3}{c}{ESC-10} & \multicolumn{3}{c}{US8K} \\
\cmidrule(lr){3-5} \cmidrule(lr){6-8} \cmidrule(lr){9-11}
Model & & 3 & 5 & 7($\sqrt{n}$) & 2 & 3($\sqrt{n}$) & 5 & 2 & 3($\sqrt{n}$) & 5 \\
\midrule
RN-50 & & 81.55 & 81.45 & \textbf{84.20} & 94.00 & 94.50 & \textbf{96.00} & 84.17 & 84.35 & \textbf{86.3} \\
EN-B0 & & 84.25 & 82.5 & \textbf{86.50} & 94.00 & 94.50 & \textbf{96.00} &  
 84.70 & \textbf{86.97} & 86.01 \\
\bottomrule
\end{tabular}
\end{table}

\subsection{Ablation Study}
\label{subsec:ablation_study}
We also perform a detailed ablation study for the impact of the size of coarse labels for the pretext task.
Table \ref{tab:ablation_study} shows the performance of various backbones across different sizes of pretext task labels where RN-50 is ResNet-50 and EN-B0 is EfficientNet-B0. We observe $\sqrt{n}$ provide
optimal performance for ESC-50\cite{Piczak2015-jy} and UrbanSound8K\cite{Salamon2014-rb} where n is the total number of different classes in the dataset. The pretext task with size as n$/$2 results in the best accuracy for ESC-10\cite{Piczak2015-jy}. We also demonstrate the visualization of the latent features using t-SNE graphs and their comparison with the baseline for the proposed framework as shown in Fig-\ref{fig:curves}.

\section{Limitation}
\label{sec:limitation}
In our proposed framework, the size of coarse labels is always less than the size of the original labels to define the label hierarchy. Therefore, our framework may not work for the task having class size $<$4 as we may not have sufficient coarse labels generated for defining the pretext task.

\section{Conclusion}
In this paper, we propose a novel framework for environment sound classification called Environmental Sound Classification with Hierarchical Ontology-guided semi-supervised Learning (ECHO) which is a two-stage coarse to fine learning framework. In the coarse learning stage, we define novel pretext tasks by generating new coarse labels based on the ontology of existing labels. We propose to use LLM-based prompt engineering to automatically determine label ontology. In the fine learning stage, we use a pre-trained model from the coarse stage and fine-tune it for the main task. On the ESC-10 dataset, our proposed approach surpasses the performance of existing works. Our experiments on various other datasets also show that the proposed framework leads to improvement of accuracy ranging from 1\% to 8\% over various baselines across four popular architectures (i.e. ResNet-18, ResNet-50, EfficientNet-B0, EfficientNet-B1). Results show that our proposed pre-text task can learn meaningful intermediate representation which helps improve the accuracy of the model on the actual task. In the future, we would like to explore the impact of our proposed framework on multi-label multi-class problems.
\bibliographystyle{unsrt}
\bibliography{my-paper.bib}

\end{document}